\documentclass[aps,reprint]{revtex4-1}

\usepackage{amssymb,amsmath}
\usepackage{graphicx}
\usepackage{caption}
\usepackage{subcaption}
\usepackage{bm}

\begin{document}


\title{Distribution of the asset price movement and market potential}


\author{Dong Han Kim}
\email{kim2010@dongguk.edu}
\thanks{Supported by the Korea Research Foundation(KRF) grant funded by the Korea government(MEST) (No. 2009-0068804).}
\affiliation{Department of Mathematics Education, Dongguk University-Seoul, 100-715 Korea}

\author{Stefano Marmi}
\email{s.marmi@sns.it}
\affiliation{Scuola Normale Superiore, Piazza dei Cavalieri 7, 56126 Pisa, Italy}
\affiliation{C.N.R.S. UMI 3483, Laboratorio Fibonacci, Piazza dei Cavalieri 7, 56126 Pisa, Italy}


\date{\today}

\begin{abstract}
In this article we discuss the distribution of asset price movements by the market potential function.
From the principle of free energy minimization we analyze two different kinds of market potentials. 
We obtain a U-shaped potential when market reversion (i.e.\ contrarian investors) is dominant.
On the other hand, if there are more trend followers, flat and logarithmic--like potentials appeared.
By using the Cyclical Adjusted Price-to-Earning ratio, which is a common valuation tool, we empirically investigate the market data.
By studying long term data we observe the historical change of the market potential of the US stock market.
Recent US data shows that the market potential looks more likely as the trending followers' potential.  
Next, we compare the market potentials for 12 different countries. 
Though some countries have similar market potentials, there are specific examples like Japan which exhibits very flat potential.
\begin{description}
\item[PACS numbers]
89.65.Gh
\end{description}

\end{abstract}


\maketitle


\section{Introduction}
In this paper we investigate the long-term behaviour of asset prices, inspired by an analogy with statistical mechanics. 
Stock market fluctuations exhibit several statistical peculiarities which are
still awaiting for a satisfactory interpretation. In a perfect, frictionless complete financial market the  movement of an asset price
should follow a random walk \cite{S65}. Real markets seem however to considerably deviate from this ideal behavior, exhibiting both significant positive serial correlation \cite{LM88} for weekly and monthly holding-period returns and long-term negative serial correlation (mean-reversion) \cite{FF88}.

In addition the price moves much faster than its real value changes \cite{S81, S00}.
According to the  efficient
markets hypothesis (EMH) the main engine of asset prices movement should be the arrival of  some new piece of information which 
leads to a revision of the expectations of market participants. 
For most assets the intrinsic value will change under the influence of slowly varying macroeconomic conditions and/or under 
the effect of unanticipated new events (discoveries, technological innovations, acquisitions, etc)

If this picture was correct, and in the absence of “noise traders”, the price should essentially be constant between two news items, and move suddenly around the release time of the news. 
Noise trading should add high frequency mean-reverting noise between news, that should not contribute to the long term volatility of the price \cite{DF}.


The price movement of an asset seems to be much more chaotic and rapid than its intrinsic value.
%
%
Although it is usually not possible to compute the present intrinsic value of an asset,
it could be estimated by statistical means after a long period of time.
The quoted price in the market moves very rapidly and it rarely equals to the intrinsic value (
indeed there is a rather strong statistical evidence that the market overpays for superior growth expectations \cite{ALS1,ALS2}).
The relation between intrinsic value and price has been well summarized by Fisher Black in 1986 \cite{B}.

In this article we analyze this phenomenon by exploiting an analogy with some ideas of statistical mechanics.
In this framework a system hardly stays at the lowest energy state (or equilibrium state). 
Indeed the system should not only minimize its energy but also maximize its entropy and the equilibrium state usually corresponds to a state with little entropy. 
Thus it generates intrinsic ``noise".
When energy is fixed the system will maximize its entropy, but if temperature is constant, it will minimize free energy. 
At high temperature, maximizing entropy is more important than minimizing energy.

Coming back to market dynamics, if we assume that the 
intrinsic value of an asset changes slowly in time and that the price of the asset should converge to the intrinsic value,
the market should exhibit some mean revertion (see, e.g.\ \cite{CF}). 
This implies the existence of a mean reverting ``force" which 
we assume to be 
a function of the deviation from the intrinsic value.
Then the force is determined by a potential function.
When the price is at the intrinsic value, it is at the lowest potential in the phase space of the asset price.
In Section~\ref{cannonical}, the detailed model for the mean reverting potential is discussed.

Whereas many long-term investors try to exploit this mean reversion there is another major investment strategy which is especially popular when the asset prices are rising: trend following. 
Following the trends means to invest only when the market is bullish and to be out of the market during a bear market.
A typical strategy for following the trends is compare the asset price with a moving average and
if the price is higher than the moving average then buy the asset and if not, go out of the market.
Others (contrarian investors) will try to exploit market reversion, so they sell the asset when its price goes up and vice versa.
We will investigate the movement of the asset price in these two different environments.

In the last Section of this paper we investigate two historical datasets on market valuations: the first
and most important has has been collected and made available\cite{Sdata} by the well-known economist Robert J. Shiller, the second has been used in reference \cite{ABMN}. 
Shiller's dataset is very widely considered by the economics and finance research communities (and also by many market practitioners)
as the reference dataset on valuation measures for the U.S. stockmarket index and has
been compiled over a very long period of time starting from
1872. To our best knowledge this is the deepest research grade historical dataset of stockmarket valuations 
which is freely available to the scientific community.

There are works on the market potential from stochastic dynamics models applied to high frequency financial time series  (e.g. \cite{BC, WTT}).
However, in this article we obtain the market potential not by making an assumption on the microscopic structure of the market dynamics 
but from the entropy maximization principle of statistical mechanics to distinguish different market behaviours on a time scale of many years.

This paper is composed as follows:
In the next section, we introduce the mathematical formulations used in this paper.
In Section~\ref{potential} we discuss various kinds of market potentials deduced by free energy minimization.
Finally, the empirical investigations of historical datasets are given in the last section.

\section{Maximum entropy principle}\label{cannonical}

The Shannon entropy of a probability density function $\rho(x)$ is defined by
$$ H (\rho) = - \int \rho(x) \log \rho(x) dx.$$
Entropy measures the uncertainty or the randomness of a random variable. The maximum entropy principle was first 
introduced by Jaynes \cite{J1, J2} in order to provide a new formulation of Statistical Mechanics based on Information Theory. 
A well-known maximum entropy principle states that when the variance of $\rho(x)$ is fixed,
the distribution which has maximum entropy is the normal distribution\cite{CT}.

Let $\varphi(x)$ be a given potential function.
Then the average energy for the density $\rho(x)$ is
$$ E = \int \varphi(x) \rho(x) dx. $$
The density function which maximizes entropy $H(\rho)$ under the condition of fixed average energy $E$ is given by
\begin{equation}\label{eq:density}
\rho_E (x) = \frac{e^{- \beta \varphi(x)}}{\int e^{- \beta \varphi(x)} dx}
\end{equation}
for some constant $\beta$ satisfying
$$\int \rho_E (x) \varphi(x) dx = \frac{\int \varphi(x) e^{- \beta \varphi(x)} dx}{\int e^{-\beta \varphi(x)} dx}  =  E.$$
(Refer to \cite{CT,NS} for the reference of the maximum entropy distributions)


Since $\log \rho_E (x) = - \beta \varphi(x) - \log \left( \int e^{-\beta \varphi(x)} dx \right)$, we have
\begin{equation*}
\begin{split}
H(\rho_E) & = - \int \rho_E \log \rho_E dx \\
&= \frac{\int \beta \varphi  e^{-\beta \varphi} dx}{\int e^{-\beta \varphi} dx} + \log \left(\int e^{-\beta \varphi} dx \right) \\
&= \beta E + \log \left( \int e^{-\beta \varphi(x)} dx \right).
\end{split}
\end{equation*}
Therefore, we have
\begin{equation*}
\begin{split}
\frac{d H(\rho_E)}{d E} 
& = \beta + E \cdot \frac{d \beta}{d E} \\
& \qquad + \frac{1}{\int e^{-\beta \varphi} dx} \cdot \frac{d}{d \beta} \left( \int e^{-\beta \varphi}dx \right) \cdot \frac{d \beta}{d E} \\
& = \beta + \frac{d \beta}{d E} \left( E - \frac{1}{\int e^{-\beta \varphi} dx} \int \varphi e^{-\beta \varphi}dx \right) \\
&= \beta.
\end{split}
\end{equation*}
Hence, we have $\beta = 1/T$ the inverse temperature.
Note that we define the entropy without Boltzmann's constant.

The density $\rho(x) = \frac{e^{- \beta \varphi(x)}}{\int e^{- \beta \varphi(x)} dx}$
maximizes the quantity
$$ H - \beta E = - \int \rho(x) \log \rho(x) dx - \beta \int \varphi(x) \rho(x) dx. $$
In other words, the density $\rho(x)$ minimizes the (Helmholtz) free energy $ F = E - T H$.



\section{Market potentials}\label{potential}

Let $S$ and $S_0$ be the price of an asset and its intrinsic value, respectively. The price $S$ changes at high frequency
whereas the intrinsic value changes in time at low frequency (apart from the rare occurrance of shocks).
Define $x = \log (S/S_0)$ so that $x=0$ corresponds to identity between price and intrinsic value (the market is perfectly 
efficient and the asset price is exactly equal to its intrinsic value). 
We assume that there is a potential $\varphi(x)$ which describes the forces which push $x$ toward the efficient equilibrium at $x=0$.

We classify the potential $\varphi(x)$ into two cases; one is the case that the force become stronger as the price deviation $|x|$ become larger
 and the other case is that the force is weaker as $|x|$ goes bigger.
We consider a system whose the force increase proportionally to $|x|$
and another system of which force decreases proportionally to $1/|x|$ for large $|x|$.

We also could use similar idea of entropy maximizing principle to obtain the distribution of the market momentum. \cite{KE}

\subsection{Market reverting potential}\label{reverting}

This is the case that there are many market practitioners who follow the mean reverting strategy.
If the price goes up, then more people sell the asset and vice versa.
The reverting force become stronger as price moves farer from the intrinsic value - the equilibrium point.
Therefore, we assume that the mean reverting force is proportional to the deviation,
so we obtain
$$\varphi(x) = C x^2$$ for some constant $C >0$.
Then we have
$$ \rho(x) = \frac{e^{- \beta x^2 }}{\int e^{- \beta x^2 } dx} ,$$
which is the Gaussian distribution.

The market temperature $\frac{d H}{d E} =  \beta = \frac{1}{T}$ is inversely corresponds to the variance of $\rho$.
In this case, the energy is
$$ E = \int x^2 \rho(x) dx$$ and $E$ corresponds the variance of the density $\rho(x)$.

\subsection{Market trending potential}\label{trends}

This is the case that there are many investors who follow the trends in the market.
If the price becomes higher, then more people come to buy the asset and vice versa.
The reverting force become weaker as price goes up or down from the intrinsic value - the equilibrium point.
Therefore, we assume that the mean reverting force is inversely proportional to the deviation. 
Thus, by integrating $1/x$, for large $|x|$ we have
$$ \varphi(x) = C \log | x | $$ for a constant $C>0$.

The density function $\rho$ is
$$\rho(x) = \frac{e^{ -\beta \log |x|}}{\int e^{- \beta \varphi(x)} dx}
= \frac{| x |^{ - \beta }}{\int e^{- \beta \varphi(x)} dx}.$$
Note that the density $\rho$ is a polynomially decreasing function
as many researches on stylized facts of market movement suggest.

The market temperature $\frac{d H}{d E} =  \beta = \frac{1}{T}$
is the exponent of $\rho(x)$.
In this case the energy is obtained as
$$ E = \frac{1}{\int e^{- \beta \varphi(x)} dx} \int | x |^{ - \beta } \log | x | dx  .$$

\subsection{Constant force}
If we assume that the mean reverting force is constant,
Then we have the density function
$$ \rho(x) = \frac{e^{- \beta |x| }}{\int e^{- \beta |x| } dx} .$$
Such potential in small time scale was considered by some articles (e.g. \cite{KC}).

\section{Empirical data}

In this section, we visualize the market potential deduced from various historical market datasets. 
In order to do this we will use 
a common valuation tool to extract the potentials from the market data.
In Section~\ref{cannonical}, we showed how the market potential function $\varphi(x)$ can be obtained from the distribution function $\rho(x)$ by \eqref{eq:density}. 
The distribution function gives the relative frequency of the deviations of the price from the equilibrium or the intrinsic value.

It is a highly challenging problem to determine the intrinsic value of financial assets.  
One of the major insights provided by the maximum entropy approach is the recommendation to focus on the behaviour of mean values of a 
rather small set of relevant quantities, instead of trying to follow the full details of the time evolution of a complex system. In a very similar 
spirit in Financial Economics the use of simple valuation tools has been proposed in order to have an approximate valuation measure 
for a market index.
Cyclically Adjusted Price-to-Earnings ratio (CAPE) is a valuation measure developed by Shiller\cite{S00}. 
It is the Price-to-Earning ratio using the average ten years of earning adjusted inflation, i.e.,
$$
\text{CAPE} = \frac{\text{price}}{\text{10 years average of inflation adjusted earning}}.
$$

Let $p_k$ be the monthly CAPE value. 
We estimate the distribution $\rho(x)$ by the relative frequency of 
the deviation of $p_k$ from the average $\bar p = \frac 1M \sum_{k=1}^M p_k$, where $M$ is the number of monthly CAPE data $p_k$.
We divide the range of $p_k - \bar p$ by equally distanced points $ a_0 < a_1 < \dots < a_n$ and set
$x_i$, $\rho_i$ to be respectively the midpoint of the $i$-th interval $[a_{i-1}, a_i)$ and of the relative frequency of $p_k - \bar p$ belonging of the $i$-th interval, i.e.,
$$x_i = \frac{a_{i-1} + a_i}{2}, \quad  \rho_i = \frac {\# \{ k : a_{i-1} < p_k - \bar p < a_i \}}{M},$$
Then, as it is discussed in Section~\ref{cannonical}, we estimate the market potential $\varphi$ at $x_i$ by 
$$ \varphi_i = - \ln \rho_i.$$

\subsection{Market potential from Historic US data}

\begin{figure*}
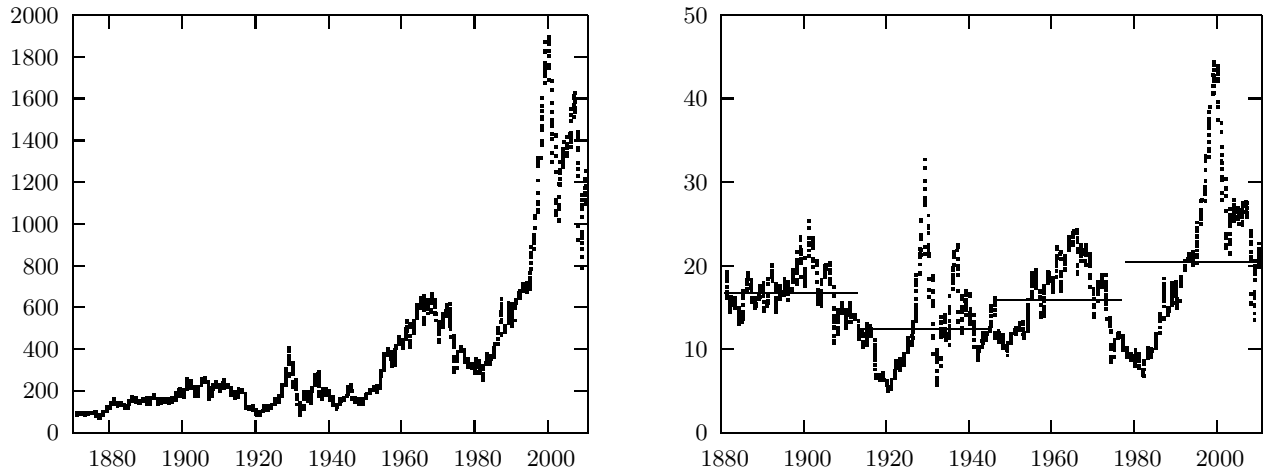

\input{SPpricegraph.tex}
\input{CAPEgraph.tex}
\caption{The historical US market date: inflation adjusted S\&P 500 stock price index from  January 1871 to February 2011 (left) and Cyclically Adjusted Price-Earnings ratio from 1881.01 to 2011.02 (CAPE) (right).
On the CAPE graph (right) the data is categorized into 4 time period, (i) 1881--1913, (ii) 1913--1945, (iii) 1946-1977, (iv) 1978--2010.
The 4 horizontal lines in CAPE graph represent the average of CAPE value in each period of time.}
\label{SPdata}\end{figure*}



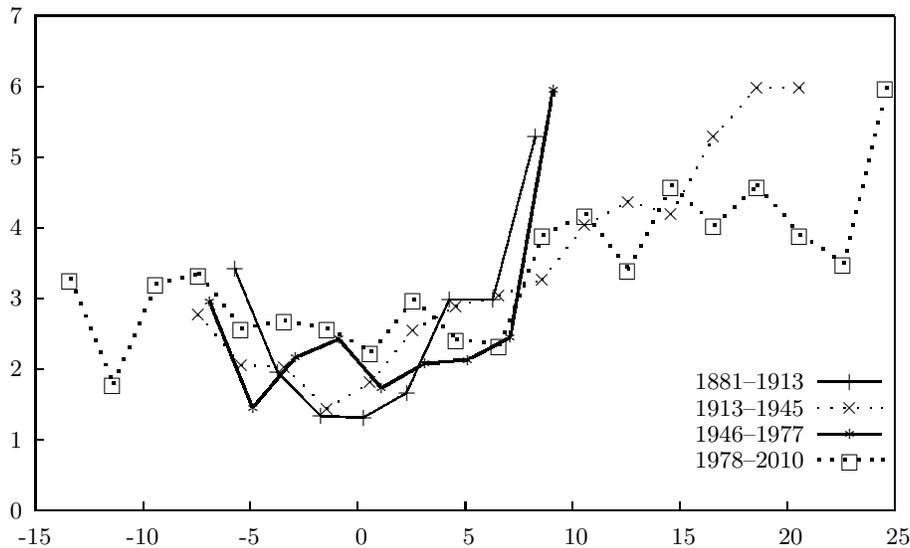
\begin{figure*}
\setlength{\unitlength}{0.240900pt}
\ifx\plotpoint\undefined\newsavebox{\plotpoint}\fi
\begin{picture}(1500,900)(0,0)
\sbox{\plotpoint}{\rule[-0.200pt]{0.400pt}{0.400pt}}%
\put(70,82){\makebox(0,0)[r]{ 0}}
\put(90.0,82.0){\rule[-0.200pt]{3.132pt}{0.400pt}}
\put(70,193){\makebox(0,0)[r]{ 1}}
\put(90.0,193.0){\rule[-0.200pt]{3.132pt}{0.400pt}}
\put(70,304){\makebox(0,0)[r]{ 2}}
\put(90.0,304.0){\rule[-0.200pt]{3.132pt}{0.400pt}}
\put(70,415){\makebox(0,0)[r]{ 3}}
\put(90.0,415.0){\rule[-0.200pt]{3.132pt}{0.400pt}}
\put(70,526){\makebox(0,0)[r]{ 4}}
\put(90.0,526.0){\rule[-0.200pt]{3.132pt}{0.400pt}}
\put(70,637){\makebox(0,0)[r]{ 5}}
\put(90.0,637.0){\rule[-0.200pt]{3.132pt}{0.400pt}}
\put(70,748){\makebox(0,0)[r]{ 6}}
\put(90.0,748.0){\rule[-0.200pt]{3.132pt}{0.400pt}}
\put(70,859){\makebox(0,0)[r]{ 7}}
\put(90.0,859.0){\rule[-0.200pt]{3.132pt}{0.400pt}}
\put(90,41){\makebox(0,0){-15}}
\put(90.0,82.0){\rule[-0.200pt]{0.400pt}{3.132pt}}
\put(259,41){\makebox(0,0){-10}}
\put(259.0,82.0){\rule[-0.200pt]{0.400pt}{3.132pt}}
\put(427,41){\makebox(0,0){-5}}
\put(427.0,82.0){\rule[-0.200pt]{0.400pt}{3.132pt}}
\put(596,41){\makebox(0,0){ 0}}
\put(596.0,82.0){\rule[-0.200pt]{0.400pt}{3.132pt}}
\put(765,41){\makebox(0,0){ 5}}
\put(765.0,82.0){\rule[-0.200pt]{0.400pt}{3.132pt}}
\put(933,41){\makebox(0,0){ 10}}
\put(933.0,82.0){\rule[-0.200pt]{0.400pt}{3.132pt}}
\put(1102,41){\makebox(0,0){ 15}}
\put(1102.0,82.0){\rule[-0.200pt]{0.400pt}{3.132pt}}
\put(1270,41){\makebox(0,0){ 20}}
\put(1270.0,82.0){\rule[-0.200pt]{0.400pt}{3.132pt}}
\put(1439,41){\makebox(0,0){ 25}}
\put(1439.0,82.0){\rule[-0.200pt]{0.400pt}{3.132pt}}
\put(90.0,82.0){\rule[-0.200pt]{0.400pt}{187.179pt}}
\put(90.0,82.0){\rule[-0.200pt]{324.974pt}{0.400pt}}
\put(1439.0,82.0){\rule[-0.200pt]{0.400pt}{187.179pt}}
\put(90.0,859.0){\rule[-0.200pt]{324.974pt}{0.400pt}}
\put(1299,284){\makebox(0,0)[r]{1881--1913}}
\put(1319.0,284.0){\rule[-0.200pt]{24.090pt}{0.400pt}}
\put(404,461){\usebox{\plotpoint}}
\multiput(404.58,456.57)(0.499,-1.212){131}{\rule{0.120pt}{1.067pt}}
\multiput(403.17,458.79)(67.000,-159.785){2}{\rule{0.400pt}{0.534pt}}
\multiput(471.58,296.90)(0.499,-0.507){133}{\rule{0.120pt}{0.506pt}}
\multiput(470.17,297.95)(68.000,-67.950){2}{\rule{0.400pt}{0.253pt}}
\put(539,228.17){\rule{13.500pt}{0.400pt}}
\multiput(539.00,229.17)(38.980,-2.000){2}{\rule{6.750pt}{0.400pt}}
\multiput(606.00,228.58)(0.874,0.498){75}{\rule{0.797pt}{0.120pt}}
\multiput(606.00,227.17)(66.345,39.000){2}{\rule{0.399pt}{0.400pt}}
\multiput(674.58,267.00)(0.499,1.092){131}{\rule{0.120pt}{0.972pt}}
\multiput(673.17,267.00)(67.000,143.983){2}{\rule{0.400pt}{0.486pt}}
\multiput(809.58,413.00)(0.499,1.917){131}{\rule{0.120pt}{1.628pt}}
\multiput(808.17,413.00)(67.000,252.620){2}{\rule{0.400pt}{0.814pt}}
\put(404,461){\makebox(0,0){$+$}}
\put(471,299){\makebox(0,0){$+$}}
\put(539,230){\makebox(0,0){$+$}}
\put(606,228){\makebox(0,0){$+$}}
\put(674,267){\makebox(0,0){$+$}}
\put(741,413){\makebox(0,0){$+$}}
\put(809,413){\makebox(0,0){$+$}}
\put(876,669){\makebox(0,0){$+$}}
\put(1369,284){\makebox(0,0){$+$}}
\put(741.0,413.0){\rule[-0.200pt]{16.381pt}{0.400pt}}
\put(1299,243){\makebox(0,0)[r]{1913--1945}}
\multiput(1319,243)(20.756,0.000){5}{\usebox{\plotpoint}}
\put(1419,243){\usebox{\plotpoint}}
\put(346,389){\usebox{\plotpoint}}
\multiput(346,389)(13.540,-15.731){6}{\usebox{\plotpoint}}
\multiput(414,310)(20.735,-0.928){3}{\usebox{\plotpoint}}
\multiput(481,307)(14.897,-14.452){4}{\usebox{\plotpoint}}
\multiput(548,242)(17.775,10.717){4}{\usebox{\plotpoint}}
\multiput(616,283)(13.132,16.073){5}{\usebox{\plotpoint}}
\multiput(683,365)(18.118,10.125){4}{\usebox{\plotpoint}}
\multiput(751,403)(20.188,4.821){3}{\usebox{\plotpoint}}
\multiput(818,419)(19.387,7.413){4}{\usebox{\plotpoint}}
\multiput(886,445)(12.849,16.300){5}{\usebox{\plotpoint}}
\multiput(953,530)(18.231,9.920){4}{\usebox{\plotpoint}}
\multiput(1021,567)(19.888,-5.937){3}{\usebox{\plotpoint}}
\multiput(1088,547)(9.991,18.193){7}{\usebox{\plotpoint}}
\multiput(1155,669)(13.739,15.557){5}{\usebox{\plotpoint}}
\multiput(1223,746)(20.756,0.000){3}{\usebox{\plotpoint}}
\put(1290,746){\usebox{\plotpoint}}
\put(346,389){\makebox(0,0){$\times$}}
\put(414,310){\makebox(0,0){$\times$}}
\put(481,307){\makebox(0,0){$\times$}}
\put(548,242){\makebox(0,0){$\times$}}
\put(616,283){\makebox(0,0){$\times$}}
\put(683,365){\makebox(0,0){$\times$}}
\put(751,403){\makebox(0,0){$\times$}}
\put(818,419){\makebox(0,0){$\times$}}
\put(886,445){\makebox(0,0){$\times$}}
\put(953,530){\makebox(0,0){$\times$}}
\put(1021,567){\makebox(0,0){$\times$}}
\put(1088,547){\makebox(0,0){$\times$}}
\put(1155,669){\makebox(0,0){$\times$}}
\put(1223,746){\makebox(0,0){$\times$}}
\put(1290,746){\makebox(0,0){$\times$}}
\put(1369,243){\makebox(0,0){$\times$}}
\sbox{\plotpoint}{\rule[-0.400pt]{0.800pt}{0.800pt}}%
\sbox{\plotpoint}{\rule[-0.200pt]{0.400pt}{0.400pt}}%
\put(1299,202){\makebox(0,0)[r]{1946--1977}}
\sbox{\plotpoint}{\rule[-0.400pt]{0.800pt}{0.800pt}}%
\put(1319.0,202.0){\rule[-0.400pt]{24.090pt}{0.800pt}}
\put(364,410){\usebox{\plotpoint}}
\multiput(365.41,401.01)(0.501,-1.234){129}{\rule{0.121pt}{2.165pt}}
\multiput(362.34,405.51)(68.000,-162.507){2}{\rule{0.800pt}{1.082pt}}
\multiput(433.41,243.00)(0.501,0.589){127}{\rule{0.121pt}{1.143pt}}
\multiput(430.34,243.00)(67.000,76.627){2}{\rule{0.800pt}{0.572pt}}
\multiput(499.00,323.41)(1.188,0.504){51}{\rule{2.076pt}{0.121pt}}
\multiput(499.00,320.34)(63.691,29.000){2}{\rule{1.038pt}{0.800pt}}
\multiput(568.41,346.35)(0.501,-0.574){127}{\rule{0.121pt}{1.119pt}}
\multiput(565.34,348.68)(67.000,-74.677){2}{\rule{0.800pt}{0.560pt}}
\multiput(634.00,275.41)(0.877,0.503){71}{\rule{1.595pt}{0.121pt}}
\multiput(634.00,272.34)(64.690,39.000){2}{\rule{0.797pt}{0.800pt}}
\multiput(702.00,314.38)(10.835,0.560){3}{\rule{10.920pt}{0.135pt}}
\multiput(702.00,311.34)(44.335,5.000){2}{\rule{5.460pt}{0.800pt}}
\multiput(769.00,319.41)(0.937,0.503){65}{\rule{1.689pt}{0.121pt}}
\multiput(769.00,316.34)(63.495,36.000){2}{\rule{0.844pt}{0.800pt}}
\multiput(837.41,354.00)(0.501,2.884){129}{\rule{0.121pt}{4.776pt}}
\multiput(834.34,354.00)(68.000,379.086){2}{\rule{0.800pt}{2.388pt}}
\put(364,410){\makebox(0,0){$\ast$}}
\put(432,243){\makebox(0,0){$\ast$}}
\put(499,322){\makebox(0,0){$\ast$}}
\put(567,351){\makebox(0,0){$\ast$}}
\put(634,274){\makebox(0,0){$\ast$}}
\put(702,313){\makebox(0,0){$\ast$}}
\put(769,318){\makebox(0,0){$\ast$}}
\put(836,354){\makebox(0,0){$\ast$}}
\put(904,743){\makebox(0,0){$\ast$}}
\put(1369,202){\makebox(0,0){$\ast$}}
\sbox{\plotpoint}{\rule[-0.500pt]{1.000pt}{1.000pt}}%
\sbox{\plotpoint}{\rule[-0.200pt]{0.400pt}{0.400pt}}%
\put(1299,161){\makebox(0,0)[r]{1978--2010}}
\sbox{\plotpoint}{\rule[-0.500pt]{1.000pt}{1.000pt}}%
\multiput(1319,161)(20.756,0.000){5}{\usebox{\plotpoint}}
\put(1419,161){\usebox{\plotpoint}}
\put(144,445){\usebox{\plotpoint}}
\multiput(144,445)(7.850,-19.214){9}{\usebox{\plotpoint}}
\multiput(211,281)(8.249,19.046){8}{\usebox{\plotpoint}}
\multiput(279,438)(20.254,4.535){4}{\usebox{\plotpoint}}
\multiput(346,453)(12.849,-16.300){5}{\usebox{\plotpoint}}
\multiput(413,368)(20.440,3.607){3}{\usebox{\plotpoint}}
\multiput(481,380)(20.430,-3.659){3}{\usebox{\plotpoint}}
\multiput(548,368)(18.231,-9.920){4}{\usebox{\plotpoint}}
\multiput(616,331)(13.132,16.073){5}{\usebox{\plotpoint}}
\multiput(683,413)(15.337,-13.984){5}{\usebox{\plotpoint}}
\multiput(751,351)(20.571,-2.763){3}{\usebox{\plotpoint}}
\multiput(818,342)(7.593,19.317){9}{\usebox{\plotpoint}}
\multiput(886,515)(18.729,8.945){3}{\usebox{\plotpoint}}
\multiput(953,547)(12.756,-16.373){6}{\usebox{\plotpoint}}
\multiput(1020,461)(9.562,18.422){7}{\usebox{\plotpoint}}
\multiput(1088,592)(15.234,-14.097){4}{\usebox{\plotpoint}}
\multiput(1155,530)(15.337,13.984){5}{\usebox{\plotpoint}}
\multiput(1223,592)(13.624,-15.658){5}{\usebox{\plotpoint}}
\multiput(1290,515)(17.309,-11.454){4}{\usebox{\plotpoint}}
\multiput(1358,470)(4.896,20.170){13}{\usebox{\plotpoint}}
\put(1425,746){\usebox{\plotpoint}}
\put(144,445){\raisebox{-.8pt}{\makebox(0,0){$\Box$}}}
\put(211,281){\raisebox{-.8pt}{\makebox(0,0){$\Box$}}}
\put(279,438){\raisebox{-.8pt}{\makebox(0,0){$\Box$}}}
\put(346,453){\raisebox{-.8pt}{\makebox(0,0){$\Box$}}}
\put(413,368){\raisebox{-.8pt}{\makebox(0,0){$\Box$}}}
\put(481,380){\raisebox{-.8pt}{\makebox(0,0){$\Box$}}}
\put(548,368){\raisebox{-.8pt}{\makebox(0,0){$\Box$}}}
\put(616,331){\raisebox{-.8pt}{\makebox(0,0){$\Box$}}}
\put(683,413){\raisebox{-.8pt}{\makebox(0,0){$\Box$}}}
\put(751,351){\raisebox{-.8pt}{\makebox(0,0){$\Box$}}}
\put(818,342){\raisebox{-.8pt}{\makebox(0,0){$\Box$}}}
\put(886,515){\raisebox{-.8pt}{\makebox(0,0){$\Box$}}}
\put(953,547){\raisebox{-.8pt}{\makebox(0,0){$\Box$}}}
\put(1020,461){\raisebox{-.8pt}{\makebox(0,0){$\Box$}}}
\put(1088,592){\raisebox{-.8pt}{\makebox(0,0){$\Box$}}}
\put(1155,530){\raisebox{-.8pt}{\makebox(0,0){$\Box$}}}
\put(1223,592){\raisebox{-.8pt}{\makebox(0,0){$\Box$}}}
\put(1290,515){\raisebox{-.8pt}{\makebox(0,0){$\Box$}}}
\put(1358,470){\raisebox{-.8pt}{\makebox(0,0){$\Box$}}}
\put(1425,746){\raisebox{-.8pt}{\makebox(0,0){$\Box$}}}
\put(1369,161){\raisebox{-.8pt}{\makebox(0,0){$\Box$}}}
\sbox{\plotpoint}{\rule[-0.200pt]{0.400pt}{0.400pt}}%
\put(90.0,82.0){\rule[-0.200pt]{0.400pt}{187.179pt}}
\put(90.0,82.0){\rule[-0.200pt]{324.974pt}{0.400pt}}
\put(1439.0,82.0){\rule[-0.200pt]{0.400pt}{187.179pt}}
\put(90.0,859.0){\rule[-0.200pt]{324.974pt}{0.400pt}}
\end{picture}
\caption{Graphs of the market potential from historical US Market data. Each graph represents the estimated potential $(x_i,\varphi_i)$ from the relative frequencies of FIG~\ref{SPdata} in one of 4 time periods (i) 1881--1913, (ii) 1913--1945, (iii) 1946-1977, (iv) 1978--2010. The abscissa of the graphs indicate the deviation from the average of CAPE (the horizontal line in the CAPE graph of FIG~\ref{SPdata})
and the ordinate is the negative logarithm of the relative frequency.}
\label{vgraph}\end{figure*}

We use the data set constructed by Shiller \cite{Sdata} to investigate the chronological changes of the market potential.
This US data set consists of monthly stock price, dividends, and earnings data and the consumer price index (to allow conversion to real values), all starting January 1871 \cite{Sdatadescription} 
(See FIG~\ref{SPdata}) .
For our purposes we will not use the dividends data. 

In FIG~\ref{vgraph}, we present how the potential in the US stock market evolves. The market data is categorized into 4 time period, (i) 1881--1913, (ii) 1913--1945, (iii) 1946-1977, (iv) 1978--2010. 
During the periods of (i) and (iii), the market potential in the US seems to be mean reverting and U-shaped as we discussed in Subsection~\ref{reverting}, thus, we may conclude that the market reversion strategy was dominant during the time span.
On the other hands, in the period of (ii) and (iv), the market potential became rather flat. It suggests that there were more market trending investors as it is considered in Subsection~\ref{trends}.

\subsection{Market potential for various countries}

\begin{figure*}
\setlength{\unitlength}{0.240900pt}
\ifx\plotpoint\undefined\newsavebox{\plotpoint}\fi
\begin{picture}(1020,779)(0,0)
\sbox{\plotpoint}{\rule[-0.200pt]{0.400pt}{0.400pt}}%
\put(70,82){\makebox(0,0)[r]{ 0}}
\put(90.0,82.0){\rule[-0.200pt]{3.132pt}{0.400pt}}
\put(70,176){\makebox(0,0)[r]{ 1}}
\put(90.0,176.0){\rule[-0.200pt]{3.132pt}{0.400pt}}
\put(70,269){\makebox(0,0)[r]{ 2}}
\put(90.0,269.0){\rule[-0.200pt]{3.132pt}{0.400pt}}
\put(70,363){\makebox(0,0)[r]{ 3}}
\put(90.0,363.0){\rule[-0.200pt]{3.132pt}{0.400pt}}
\put(70,457){\makebox(0,0)[r]{ 4}}
\put(90.0,457.0){\rule[-0.200pt]{3.132pt}{0.400pt}}
\put(70,551){\makebox(0,0)[r]{ 5}}
\put(90.0,551.0){\rule[-0.200pt]{3.132pt}{0.400pt}}
\put(70,644){\makebox(0,0)[r]{ 6}}
\put(90.0,644.0){\rule[-0.200pt]{3.132pt}{0.400pt}}
\put(70,738){\makebox(0,0)[r]{ 7}}
\put(90.0,738.0){\rule[-0.200pt]{3.132pt}{0.400pt}}
\put(90,41){\makebox(0,0){-20}}
\put(90.0,82.0){\rule[-0.200pt]{0.400pt}{3.132pt}}
\put(228,41){\makebox(0,0){-10}}
\put(228.0,82.0){\rule[-0.200pt]{0.400pt}{3.132pt}}
\put(366,41){\makebox(0,0){ 0}}
\put(366.0,82.0){\rule[-0.200pt]{0.400pt}{3.132pt}}
\put(504,41){\makebox(0,0){ 10}}
\put(504.0,82.0){\rule[-0.200pt]{0.400pt}{3.132pt}}
\put(642,41){\makebox(0,0){ 20}}
\put(642.0,82.0){\rule[-0.200pt]{0.400pt}{3.132pt}}
\put(780,41){\makebox(0,0){ 30}}
\put(780.0,82.0){\rule[-0.200pt]{0.400pt}{3.132pt}}
\put(918,41){\makebox(0,0){ 40}}
\put(918.0,82.0){\rule[-0.200pt]{0.400pt}{3.132pt}}
\put(90.0,82.0){\rule[-0.200pt]{0.400pt}{158.030pt}}
\put(90.0,82.0){\rule[-0.200pt]{209.342pt}{0.400pt}}
\put(959.0,82.0){\rule[-0.200pt]{0.400pt}{158.030pt}}
\put(90.0,738.0){\rule[-0.200pt]{209.342pt}{0.400pt}}
\put(819,249){\makebox(0,0)[r]{Norway}}
\put(839.0,249.0){\rule[-0.200pt]{24.090pt}{0.400pt}}
\put(217,501){\usebox{\plotpoint}}
\multiput(217.58,488.44)(0.497,-3.691){61}{\rule{0.120pt}{3.025pt}}
\multiput(216.17,494.72)(32.000,-227.721){2}{\rule{0.400pt}{1.513pt}}
\multiput(249.00,265.92)(0.528,-0.498){101}{\rule{0.523pt}{0.120pt}}
\multiput(249.00,266.17)(53.914,-52.000){2}{\rule{0.262pt}{0.400pt}}
\multiput(304.00,213.92)(2.893,-0.491){17}{\rule{2.340pt}{0.118pt}}
\multiput(304.00,214.17)(51.143,-10.000){2}{\rule{1.170pt}{0.400pt}}
\multiput(360.00,205.58)(0.573,0.498){93}{\rule{0.558pt}{0.120pt}}
\multiput(360.00,204.17)(53.841,48.000){2}{\rule{0.279pt}{0.400pt}}
\multiput(415.00,253.58)(0.550,0.498){97}{\rule{0.540pt}{0.120pt}}
\multiput(415.00,252.17)(53.879,50.000){2}{\rule{0.270pt}{0.400pt}}
\multiput(470.58,303.00)(0.499,0.792){107}{\rule{0.120pt}{0.733pt}}
\multiput(469.17,303.00)(55.000,85.479){2}{\rule{0.400pt}{0.366pt}}
\multiput(580.58,390.00)(0.499,1.236){109}{\rule{0.120pt}{1.086pt}}
\multiput(579.17,390.00)(56.000,135.747){2}{\rule{0.400pt}{0.543pt}}
\put(217,501){\makebox(0,0){$+$}}
\put(249,267){\makebox(0,0){$+$}}
\put(304,215){\makebox(0,0){$+$}}
\put(360,205){\makebox(0,0){$+$}}
\put(415,253){\makebox(0,0){$+$}}
\put(470,303){\makebox(0,0){$+$}}
\put(525,390){\makebox(0,0){$+$}}
\put(580,390){\makebox(0,0){$+$}}
\put(636,528){\makebox(0,0){$+$}}
\put(889,249){\makebox(0,0){$+$}}
\put(525.0,390.0){\rule[-0.200pt]{13.249pt}{0.400pt}}
\put(819,208){\makebox(0,0)[r]{UK}}
\multiput(839,208)(20.756,0.000){5}{\usebox{\plotpoint}}
\put(939,208){\usebox{\plotpoint}}
\put(232,463){\usebox{\plotpoint}}
\multiput(232,463)(3.115,-20.520){11}{\usebox{\plotpoint}}
\multiput(266,239)(14.153,15.182){4}{\usebox{\plotpoint}}
\multiput(321,298)(8.035,-19.137){7}{\usebox{\plotpoint}}
\multiput(376,167)(13.169,16.043){4}{\usebox{\plotpoint}}
\multiput(431,234)(13.079,16.116){5}{\usebox{\plotpoint}}
\put(487,303){\usebox{\plotpoint}}
\put(232,463){\makebox(0,0){$\times$}}
\put(266,239){\makebox(0,0){$\times$}}
\put(321,298){\makebox(0,0){$\times$}}
\put(376,167){\makebox(0,0){$\times$}}
\put(431,234){\makebox(0,0){$\times$}}
\put(487,303){\makebox(0,0){$\times$}}
\put(889,208){\makebox(0,0){$\times$}}
\sbox{\plotpoint}{\rule[-0.400pt]{0.800pt}{0.800pt}}%
\sbox{\plotpoint}{\rule[-0.200pt]{0.400pt}{0.400pt}}%
\put(819,167){\makebox(0,0)[r]{Belguim}}
\sbox{\plotpoint}{\rule[-0.400pt]{0.800pt}{0.800pt}}%
\put(839.0,167.0){\rule[-0.400pt]{24.090pt}{0.800pt}}
\put(212,390){\usebox{\plotpoint}}
\multiput(213.41,376.41)(0.503,-1.954){57}{\rule{0.121pt}{3.275pt}}
\multiput(210.34,383.20)(32.000,-116.203){2}{\rule{0.800pt}{1.638pt}}
\multiput(244.00,268.41)(0.656,0.502){77}{\rule{1.248pt}{0.121pt}}
\multiput(244.00,265.34)(52.411,42.000){2}{\rule{0.624pt}{0.800pt}}
\multiput(300.41,301.35)(0.502,-1.033){103}{\rule{0.121pt}{1.844pt}}
\multiput(297.34,305.17)(55.000,-109.173){2}{\rule{0.800pt}{0.922pt}}
\multiput(354.00,197.41)(1.337,0.505){35}{\rule{2.295pt}{0.122pt}}
\multiput(354.00,194.34)(50.236,21.000){2}{\rule{1.148pt}{0.800pt}}
\multiput(409.00,218.41)(1.031,0.504){47}{\rule{1.830pt}{0.121pt}}
\multiput(409.00,215.34)(51.203,27.000){2}{\rule{0.915pt}{0.800pt}}
\multiput(465.41,244.00)(0.502,1.837){105}{\rule{0.121pt}{3.114pt}}
\multiput(462.34,244.00)(56.000,197.536){2}{\rule{0.800pt}{1.557pt}}
\multiput(521.41,448.00)(0.502,1.079){103}{\rule{0.121pt}{1.916pt}}
\multiput(518.34,448.00)(55.000,114.022){2}{\rule{0.800pt}{0.958pt}}
\put(212,390){\makebox(0,0){$\ast$}}
\put(244,267){\makebox(0,0){$\ast$}}
\put(299,309){\makebox(0,0){$\ast$}}
\put(354,196){\makebox(0,0){$\ast$}}
\put(409,217){\makebox(0,0){$\ast$}}
\put(464,244){\makebox(0,0){$\ast$}}
\put(520,448){\makebox(0,0){$\ast$}}
\put(575,566){\makebox(0,0){$\ast$}}
\put(889,167){\makebox(0,0){$\ast$}}
\sbox{\plotpoint}{\rule[-0.500pt]{1.000pt}{1.000pt}}%
\sbox{\plotpoint}{\rule[-0.200pt]{0.400pt}{0.400pt}}%
\put(819,126){\makebox(0,0)[r]{Netherlands}}
\sbox{\plotpoint}{\rule[-0.500pt]{1.000pt}{1.000pt}}%
\multiput(839,126)(20.756,0.000){5}{\usebox{\plotpoint}}
\put(939,126){\usebox{\plotpoint}}
\put(220,326){\usebox{\plotpoint}}
\multiput(220,326)(10.441,-17.938){4}{\usebox{\plotpoint}}
\multiput(259,259)(20.701,-1.506){3}{\usebox{\plotpoint}}
\multiput(314,255)(14.282,-15.061){4}{\usebox{\plotpoint}}
\multiput(369,197)(13.194,16.022){4}{\usebox{\plotpoint}}
\multiput(425,265)(12.599,16.494){4}{\usebox{\plotpoint}}
\multiput(480,337)(16.496,-12.597){4}{\usebox{\plotpoint}}
\multiput(535,295)(5.915,19.895){9}{\usebox{\plotpoint}}
\put(590,480){\usebox{\plotpoint}}
\put(220,326){\raisebox{-.8pt}{\makebox(0,0){$\Box$}}}
\put(259,259){\raisebox{-.8pt}{\makebox(0,0){$\Box$}}}
\put(314,255){\raisebox{-.8pt}{\makebox(0,0){$\Box$}}}
\put(369,197){\raisebox{-.8pt}{\makebox(0,0){$\Box$}}}
\put(425,265){\raisebox{-.8pt}{\makebox(0,0){$\Box$}}}
\put(480,337){\raisebox{-.8pt}{\makebox(0,0){$\Box$}}}
\put(535,295){\raisebox{-.8pt}{\makebox(0,0){$\Box$}}}
\put(590,480){\raisebox{-.8pt}{\makebox(0,0){$\Box$}}}
\put(889,126){\raisebox{-.8pt}{\makebox(0,0){$\Box$}}}
\sbox{\plotpoint}{\rule[-0.200pt]{0.400pt}{0.400pt}}%
\put(90.0,82.0){\rule[-0.200pt]{0.400pt}{158.030pt}}
\put(90.0,82.0){\rule[-0.200pt]{209.342pt}{0.400pt}}
\put(959.0,82.0){\rule[-0.200pt]{0.400pt}{158.030pt}}
\put(90.0,738.0){\rule[-0.200pt]{209.342pt}{0.400pt}}
\end{picture}
\setlength{\unitlength}{0.240900pt}
\ifx\plotpoint\undefined\newsavebox{\plotpoint}\fi
\begin{picture}(1020,779)(0,0)
\sbox{\plotpoint}{\rule[-0.200pt]{0.400pt}{0.400pt}}%
\put(70,82){\makebox(0,0)[r]{ 0}}
\put(90.0,82.0){\rule[-0.200pt]{3.132pt}{0.400pt}}
\put(70,176){\makebox(0,0)[r]{ 1}}
\put(90.0,176.0){\rule[-0.200pt]{3.132pt}{0.400pt}}
\put(70,269){\makebox(0,0)[r]{ 2}}
\put(90.0,269.0){\rule[-0.200pt]{3.132pt}{0.400pt}}
\put(70,363){\makebox(0,0)[r]{ 3}}
\put(90.0,363.0){\rule[-0.200pt]{3.132pt}{0.400pt}}
\put(70,457){\makebox(0,0)[r]{ 4}}
\put(90.0,457.0){\rule[-0.200pt]{3.132pt}{0.400pt}}
\put(70,551){\makebox(0,0)[r]{ 5}}
\put(90.0,551.0){\rule[-0.200pt]{3.132pt}{0.400pt}}
\put(70,644){\makebox(0,0)[r]{ 6}}
\put(90.0,644.0){\rule[-0.200pt]{3.132pt}{0.400pt}}
\put(70,738){\makebox(0,0)[r]{ 7}}
\put(90.0,738.0){\rule[-0.200pt]{3.132pt}{0.400pt}}
\put(90,41){\makebox(0,0){-20}}
\put(90.0,82.0){\rule[-0.200pt]{0.400pt}{3.132pt}}
\put(228,41){\makebox(0,0){-10}}
\put(228.0,82.0){\rule[-0.200pt]{0.400pt}{3.132pt}}
\put(366,41){\makebox(0,0){ 0}}
\put(366.0,82.0){\rule[-0.200pt]{0.400pt}{3.132pt}}
\put(504,41){\makebox(0,0){ 10}}
\put(504.0,82.0){\rule[-0.200pt]{0.400pt}{3.132pt}}
\put(642,41){\makebox(0,0){ 20}}
\put(642.0,82.0){\rule[-0.200pt]{0.400pt}{3.132pt}}
\put(780,41){\makebox(0,0){ 30}}
\put(780.0,82.0){\rule[-0.200pt]{0.400pt}{3.132pt}}
\put(918,41){\makebox(0,0){ 40}}
\put(918.0,82.0){\rule[-0.200pt]{0.400pt}{3.132pt}}
\put(90.0,82.0){\rule[-0.200pt]{0.400pt}{158.030pt}}
\put(90.0,82.0){\rule[-0.200pt]{209.342pt}{0.400pt}}
\put(959.0,82.0){\rule[-0.200pt]{0.400pt}{158.030pt}}
\put(90.0,738.0){\rule[-0.200pt]{209.342pt}{0.400pt}}
\put(819,249){\makebox(0,0)[r]{USA}}
\put(839.0,249.0){\rule[-0.200pt]{24.090pt}{0.400pt}}
\put(193,246){\usebox{\plotpoint}}
\multiput(193.58,246.00)(0.499,0.508){109}{\rule{0.120pt}{0.507pt}}
\multiput(192.17,246.00)(56.000,55.947){2}{\rule{0.400pt}{0.254pt}}
\multiput(249.00,301.92)(0.598,-0.498){89}{\rule{0.578pt}{0.120pt}}
\multiput(249.00,302.17)(53.800,-46.000){2}{\rule{0.289pt}{0.400pt}}
\put(304,255.17){\rule{11.100pt}{0.400pt}}
\multiput(304.00,256.17)(31.961,-2.000){2}{\rule{5.550pt}{0.400pt}}
\multiput(359.00,253.92)(0.862,-0.497){61}{\rule{0.788pt}{0.120pt}}
\multiput(359.00,254.17)(53.366,-32.000){2}{\rule{0.394pt}{0.400pt}}
\multiput(414.58,223.00)(0.499,1.250){107}{\rule{0.120pt}{1.096pt}}
\multiput(413.17,223.00)(55.000,134.724){2}{\rule{0.400pt}{0.548pt}}
\multiput(469.00,360.59)(6.165,0.477){7}{\rule{4.580pt}{0.115pt}}
\multiput(469.00,359.17)(46.494,5.000){2}{\rule{2.290pt}{0.400pt}}
\multiput(525.00,365.58)(0.539,0.498){99}{\rule{0.531pt}{0.120pt}}
\multiput(525.00,364.17)(53.897,51.000){2}{\rule{0.266pt}{0.400pt}}
\multiput(580.00,414.92)(0.625,-0.498){85}{\rule{0.600pt}{0.120pt}}
\multiput(580.00,415.17)(53.755,-44.000){2}{\rule{0.300pt}{0.400pt}}
\multiput(635.58,372.00)(0.499,0.582){107}{\rule{0.120pt}{0.565pt}}
\multiput(634.17,372.00)(55.000,62.826){2}{\rule{0.400pt}{0.283pt}}
\put(193,246){\makebox(0,0){$+$}}
\put(249,303){\makebox(0,0){$+$}}
\put(304,257){\makebox(0,0){$+$}}
\put(359,255){\makebox(0,0){$+$}}
\put(414,223){\makebox(0,0){$+$}}
\put(469,360){\makebox(0,0){$+$}}
\put(525,365){\makebox(0,0){$+$}}
\put(580,416){\makebox(0,0){$+$}}
\put(635,372){\makebox(0,0){$+$}}
\put(690,436){\makebox(0,0){$+$}}
\put(889,249){\makebox(0,0){$+$}}
\put(819,208){\makebox(0,0)[r]{Germany}}
\multiput(839,208)(20.756,0.000){5}{\usebox{\plotpoint}}
\put(939,208){\usebox{\plotpoint}}
\put(167,268){\usebox{\plotpoint}}
\multiput(167,268)(12.381,16.658){5}{\usebox{\plotpoint}}
\multiput(222,342)(16.931,-12.006){3}{\usebox{\plotpoint}}
\multiput(277,303)(12.823,-16.320){4}{\usebox{\plotpoint}}
\multiput(332,233)(20.637,2.211){3}{\usebox{\plotpoint}}
\multiput(388,239)(18.497,9.416){3}{\usebox{\plotpoint}}
\multiput(443,267)(11.961,16.963){5}{\usebox{\plotpoint}}
\multiput(498,345)(19.618,-6.777){3}{\usebox{\plotpoint}}
\multiput(553,326)(10.823,17.710){5}{\usebox{\plotpoint}}
\multiput(608,416)(13.668,15.620){4}{\usebox{\plotpoint}}
\multiput(664,480)(7.103,19.502){7}{\usebox{\plotpoint}}
\put(719,631){\usebox{\plotpoint}}
\put(167,268){\makebox(0,0){$\times$}}
\put(222,342){\makebox(0,0){$\times$}}
\put(277,303){\makebox(0,0){$\times$}}
\put(332,233){\makebox(0,0){$\times$}}
\put(388,239){\makebox(0,0){$\times$}}
\put(443,267){\makebox(0,0){$\times$}}
\put(498,345){\makebox(0,0){$\times$}}
\put(553,326){\makebox(0,0){$\times$}}
\put(608,416){\makebox(0,0){$\times$}}
\put(664,480){\makebox(0,0){$\times$}}
\put(719,631){\makebox(0,0){$\times$}}
\put(889,208){\makebox(0,0){$\times$}}
\sbox{\plotpoint}{\rule[-0.400pt]{0.800pt}{0.800pt}}%
\sbox{\plotpoint}{\rule[-0.200pt]{0.400pt}{0.400pt}}%
\put(819,167){\makebox(0,0)[r]{Australia}}
\sbox{\plotpoint}{\rule[-0.400pt]{0.800pt}{0.800pt}}%
\put(839.0,167.0){\rule[-0.400pt]{24.090pt}{0.800pt}}
\put(248,252){\usebox{\plotpoint}}
\multiput(248.00,250.09)(1.274,-0.505){37}{\rule{2.200pt}{0.122pt}}
\multiput(248.00,250.34)(50.434,-22.000){2}{\rule{1.100pt}{0.800pt}}
\multiput(303.00,228.09)(0.549,-0.502){93}{\rule{1.080pt}{0.121pt}}
\multiput(303.00,228.34)(52.758,-50.000){2}{\rule{0.540pt}{0.800pt}}
\multiput(359.41,180.00)(0.502,0.821){103}{\rule{0.121pt}{1.509pt}}
\multiput(356.34,180.00)(55.000,86.868){2}{\rule{0.800pt}{0.755pt}}
\multiput(414.41,270.00)(0.502,1.632){103}{\rule{0.121pt}{2.789pt}}
\multiput(411.34,270.00)(55.000,172.211){2}{\rule{0.800pt}{1.395pt}}
\multiput(469.41,440.59)(0.502,-0.996){105}{\rule{0.121pt}{1.786pt}}
\multiput(466.34,444.29)(56.000,-107.294){2}{\rule{0.800pt}{0.893pt}}
\multiput(525.41,337.00)(0.502,0.554){103}{\rule{0.121pt}{1.087pt}}
\multiput(522.34,337.00)(55.000,58.743){2}{\rule{0.800pt}{0.544pt}}
\multiput(579.00,399.41)(0.726,0.503){69}{\rule{1.358pt}{0.121pt}}
\multiput(579.00,396.34)(52.182,38.000){2}{\rule{0.679pt}{0.800pt}}
\multiput(635.41,436.00)(0.502,1.189){103}{\rule{0.121pt}{2.091pt}}
\multiput(632.34,436.00)(55.000,125.660){2}{\rule{0.800pt}{1.045pt}}
\put(248,252){\makebox(0,0){$\ast$}}
\put(303,230){\makebox(0,0){$\ast$}}
\put(358,180){\makebox(0,0){$\ast$}}
\put(413,270){\makebox(0,0){$\ast$}}
\put(468,448){\makebox(0,0){$\ast$}}
\put(524,337){\makebox(0,0){$\ast$}}
\put(579,398){\makebox(0,0){$\ast$}}
\put(634,436){\makebox(0,0){$\ast$}}
\put(689,566){\makebox(0,0){$\ast$}}
\put(889,167){\makebox(0,0){$\ast$}}
\sbox{\plotpoint}{\rule[-0.200pt]{0.400pt}{0.400pt}}%
\put(90.0,82.0){\rule[-0.200pt]{0.400pt}{158.030pt}}
\put(90.0,82.0){\rule[-0.200pt]{209.342pt}{0.400pt}}
\put(959.0,82.0){\rule[-0.200pt]{0.400pt}{158.030pt}}
\put(90.0,738.0){\rule[-0.200pt]{209.342pt}{0.400pt}}
\end{picture}

\input{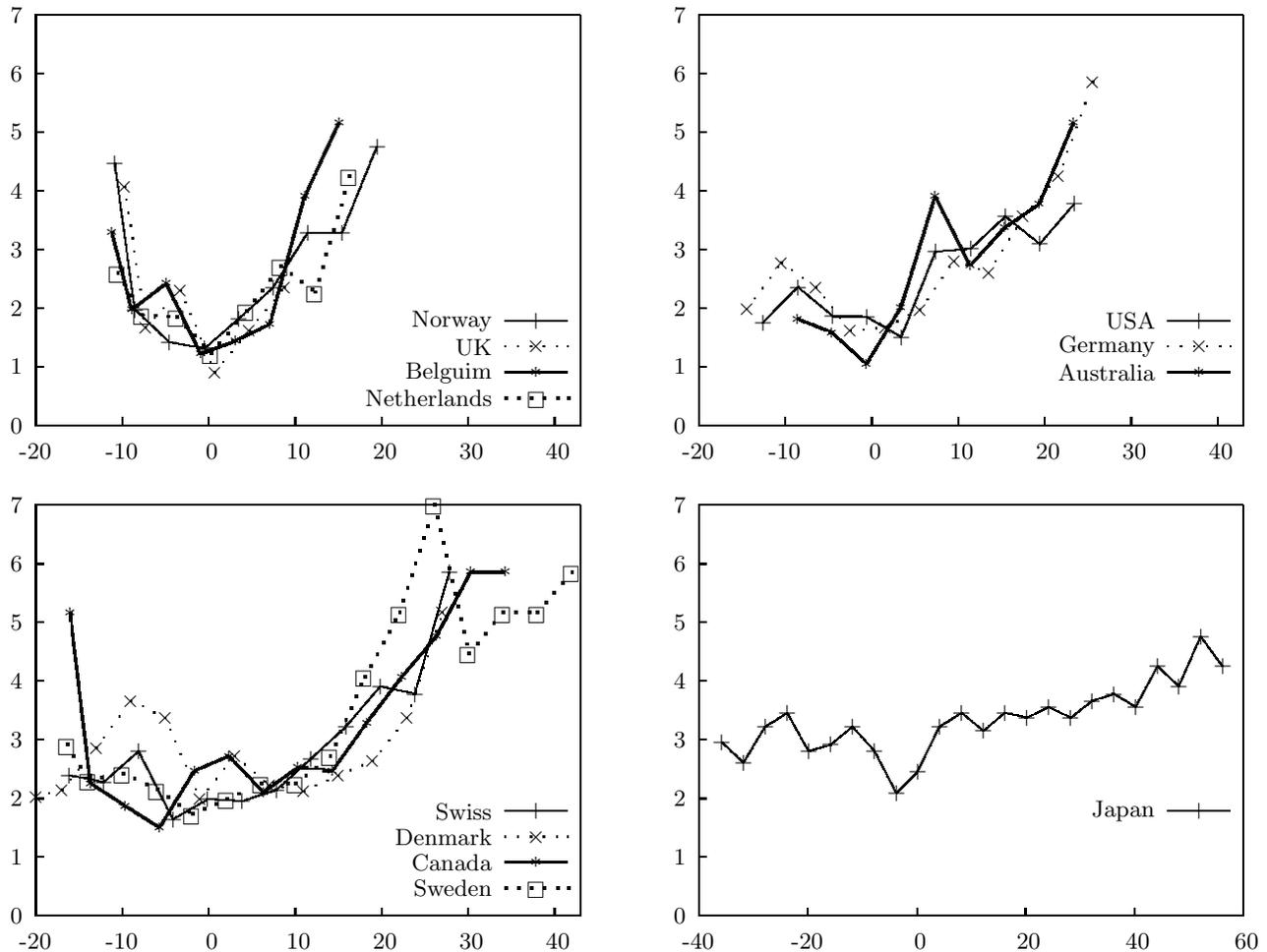}
\setlength{\unitlength}{0.240900pt}
\ifx\plotpoint\undefined\newsavebox{\plotpoint}\fi
\begin{picture}(1020,779)(0,0)
\sbox{\plotpoint}{\rule[-0.200pt]{0.400pt}{0.400pt}}%
\put(70,82){\makebox(0,0)[r]{ 0}}
\put(90.0,82.0){\rule[-0.200pt]{3.132pt}{0.400pt}}
\put(70,176){\makebox(0,0)[r]{ 1}}
\put(90.0,176.0){\rule[-0.200pt]{3.132pt}{0.400pt}}
\put(70,269){\makebox(0,0)[r]{ 2}}
\put(90.0,269.0){\rule[-0.200pt]{3.132pt}{0.400pt}}
\put(70,363){\makebox(0,0)[r]{ 3}}
\put(90.0,363.0){\rule[-0.200pt]{3.132pt}{0.400pt}}
\put(70,457){\makebox(0,0)[r]{ 4}}
\put(90.0,457.0){\rule[-0.200pt]{3.132pt}{0.400pt}}
\put(70,551){\makebox(0,0)[r]{ 5}}
\put(90.0,551.0){\rule[-0.200pt]{3.132pt}{0.400pt}}
\put(70,644){\makebox(0,0)[r]{ 6}}
\put(90.0,644.0){\rule[-0.200pt]{3.132pt}{0.400pt}}
\put(70,738){\makebox(0,0)[r]{ 7}}
\put(90.0,738.0){\rule[-0.200pt]{3.132pt}{0.400pt}}
\put(90,41){\makebox(0,0){-40}}
\put(90.0,82.0){\rule[-0.200pt]{0.400pt}{3.132pt}}
\put(264,41){\makebox(0,0){-20}}
\put(264.0,82.0){\rule[-0.200pt]{0.400pt}{3.132pt}}
\put(438,41){\makebox(0,0){ 0}}
\put(438.0,82.0){\rule[-0.200pt]{0.400pt}{3.132pt}}
\put(611,41){\makebox(0,0){ 20}}
\put(611.0,82.0){\rule[-0.200pt]{0.400pt}{3.132pt}}
\put(785,41){\makebox(0,0){ 40}}
\put(785.0,82.0){\rule[-0.200pt]{0.400pt}{3.132pt}}
\put(959,41){\makebox(0,0){ 60}}
\put(959.0,82.0){\rule[-0.200pt]{0.400pt}{3.132pt}}
\put(90.0,82.0){\rule[-0.200pt]{0.400pt}{158.030pt}}
\put(90.0,82.0){\rule[-0.200pt]{209.342pt}{0.400pt}}
\put(959.0,82.0){\rule[-0.200pt]{0.400pt}{158.030pt}}
\put(90.0,738.0){\rule[-0.200pt]{209.342pt}{0.400pt}}
\put(819,249){\makebox(0,0)[r]{Japan}}
\put(839.0,249.0){\rule[-0.200pt]{24.090pt}{0.400pt}}
\put(127,359){\usebox{\plotpoint}}
\multiput(127.00,357.92)(0.529,-0.497){63}{\rule{0.524pt}{0.120pt}}
\multiput(127.00,358.17)(33.912,-33.000){2}{\rule{0.262pt}{0.400pt}}
\multiput(162.58,326.00)(0.498,0.831){67}{\rule{0.120pt}{0.763pt}}
\multiput(161.17,326.00)(35.000,56.417){2}{\rule{0.400pt}{0.381pt}}
\multiput(197.00,384.58)(0.799,0.496){41}{\rule{0.736pt}{0.120pt}}
\multiput(197.00,383.17)(33.472,22.000){2}{\rule{0.368pt}{0.400pt}}
\multiput(232.58,402.61)(0.498,-0.900){65}{\rule{0.120pt}{0.818pt}}
\multiput(231.17,404.30)(34.000,-59.303){2}{\rule{0.400pt}{0.409pt}}
\multiput(266.00,345.58)(1.798,0.491){17}{\rule{1.500pt}{0.118pt}}
\multiput(266.00,344.17)(31.887,10.000){2}{\rule{0.750pt}{0.400pt}}
\multiput(301.00,355.58)(0.603,0.497){55}{\rule{0.583pt}{0.120pt}}
\multiput(301.00,354.17)(33.790,29.000){2}{\rule{0.291pt}{0.400pt}}
\multiput(336.58,381.73)(0.498,-0.557){67}{\rule{0.120pt}{0.546pt}}
\multiput(335.17,382.87)(35.000,-37.867){2}{\rule{0.400pt}{0.273pt}}
\multiput(371.58,341.41)(0.498,-0.961){67}{\rule{0.120pt}{0.866pt}}
\multiput(370.17,343.20)(35.000,-65.203){2}{\rule{0.400pt}{0.433pt}}
\multiput(406.00,278.58)(0.499,0.498){65}{\rule{0.500pt}{0.120pt}}
\multiput(406.00,277.17)(32.962,34.000){2}{\rule{0.250pt}{0.400pt}}
\multiput(440.58,312.00)(0.498,1.033){67}{\rule{0.120pt}{0.923pt}}
\multiput(439.17,312.00)(35.000,70.085){2}{\rule{0.400pt}{0.461pt}}
\multiput(475.00,384.58)(0.799,0.496){41}{\rule{0.736pt}{0.120pt}}
\multiput(475.00,383.17)(33.472,22.000){2}{\rule{0.368pt}{0.400pt}}
\multiput(510.00,404.92)(0.603,-0.497){55}{\rule{0.583pt}{0.120pt}}
\multiput(510.00,405.17)(33.790,-29.000){2}{\rule{0.291pt}{0.400pt}}
\multiput(545.00,377.58)(0.586,0.497){55}{\rule{0.569pt}{0.120pt}}
\multiput(545.00,376.17)(32.819,29.000){2}{\rule{0.284pt}{0.400pt}}
\multiput(579.00,404.93)(2.277,-0.488){13}{\rule{1.850pt}{0.117pt}}
\multiput(579.00,405.17)(31.160,-8.000){2}{\rule{0.925pt}{0.400pt}}
\multiput(614.00,398.58)(1.039,0.495){31}{\rule{0.924pt}{0.119pt}}
\multiput(614.00,397.17)(33.083,17.000){2}{\rule{0.462pt}{0.400pt}}
\multiput(649.00,413.92)(1.039,-0.495){31}{\rule{0.924pt}{0.119pt}}
\multiput(649.00,414.17)(33.083,-17.000){2}{\rule{0.462pt}{0.400pt}}
\multiput(684.00,398.58)(0.630,0.497){51}{\rule{0.604pt}{0.120pt}}
\multiput(684.00,397.17)(32.747,27.000){2}{\rule{0.302pt}{0.400pt}}
\multiput(718.00,425.58)(1.628,0.492){19}{\rule{1.373pt}{0.118pt}}
\multiput(718.00,424.17)(32.151,11.000){2}{\rule{0.686pt}{0.400pt}}
\multiput(753.00,434.92)(0.837,-0.496){39}{\rule{0.767pt}{0.119pt}}
\multiput(753.00,435.17)(33.409,-21.000){2}{\rule{0.383pt}{0.400pt}}
\multiput(788.58,415.00)(0.498,0.932){67}{\rule{0.120pt}{0.843pt}}
\multiput(787.17,415.00)(35.000,63.251){2}{\rule{0.400pt}{0.421pt}}
\multiput(823.00,478.92)(0.530,-0.497){61}{\rule{0.525pt}{0.120pt}}
\multiput(823.00,479.17)(32.910,-32.000){2}{\rule{0.263pt}{0.400pt}}
\multiput(857.58,448.00)(0.498,1.148){67}{\rule{0.120pt}{1.014pt}}
\multiput(856.17,448.00)(35.000,77.895){2}{\rule{0.400pt}{0.507pt}}
\multiput(892.58,525.31)(0.498,-0.686){67}{\rule{0.120pt}{0.649pt}}
\multiput(891.17,526.65)(35.000,-46.654){2}{\rule{0.400pt}{0.324pt}}
\put(127,359){\makebox(0,0){$+$}}
\put(162,326){\makebox(0,0){$+$}}
\put(197,384){\makebox(0,0){$+$}}
\put(232,406){\makebox(0,0){$+$}}
\put(266,345){\makebox(0,0){$+$}}
\put(301,355){\makebox(0,0){$+$}}
\put(336,384){\makebox(0,0){$+$}}
\put(371,345){\makebox(0,0){$+$}}
\put(406,278){\makebox(0,0){$+$}}
\put(440,312){\makebox(0,0){$+$}}
\put(475,384){\makebox(0,0){$+$}}
\put(510,406){\makebox(0,0){$+$}}
\put(545,377){\makebox(0,0){$+$}}
\put(579,406){\makebox(0,0){$+$}}
\put(614,398){\makebox(0,0){$+$}}
\put(649,415){\makebox(0,0){$+$}}
\put(684,398){\makebox(0,0){$+$}}
\put(718,425){\makebox(0,0){$+$}}
\put(753,436){\makebox(0,0){$+$}}
\put(788,415){\makebox(0,0){$+$}}
\put(823,480){\makebox(0,0){$+$}}
\put(857,448){\makebox(0,0){$+$}}
\put(892,528){\makebox(0,0){$+$}}
\put(927,480){\makebox(0,0){$+$}}
\put(889,249){\makebox(0,0){$+$}}
\put(90.0,82.0){\rule[-0.200pt]{0.400pt}{158.030pt}}
\put(90.0,82.0){\rule[-0.200pt]{209.342pt}{0.400pt}}
\put(959.0,82.0){\rule[-0.200pt]{0.400pt}{158.030pt}}
\put(90.0,738.0){\rule[-0.200pt]{209.342pt}{0.400pt}}
\end{picture}
\caption{Graphs of the market potential from various countries - Australia, Belgium, Canada, Denmark, Germany, Japan, the Netherlands, Norway, Sweden, Switzerland, and United Kingdom by recent data. Each graph represents the estimated potential $(x_i,\varphi_i)$ from the relative frequency. The abscissa indicate the deviation from the historical average of CAPE and the ordinate is the negative logarithm of the relative frequency as before.}
\label{hgraph}\end{figure*}

For Australia, Belgium, Canada, Denmark, Germany, Japan, the Netherlands, Norway, Sweden, Switzerland, 
and United Kingdom we use the CAPE dataset \cite{ABMNdata} built in \cite{ABMN}. 
At variance with US dataset, these series span a narrower time window running from December 1969 to December 2010. 
The estimated market potential $(x_i,\varphi_i)$ for these 12 countries are shown in FIG~\ref{hgraph}.
Among them Norway, UK, Belgium and Netherlands can be categorized as countries with market reverting potentials. 
Other countries like USA, Switzerland, Germany and Denmark have similar market potential. 
The most dramatic example is observed in Japan, which has a very flat potential.


\begin{thebibliography}{99}


\bibitem{ABMN}
N. Angelini, G. Bormetti, S. Marmi and F. Nardini, {\em Value Matters: Predictability of stock index returns}, preprint.

\bibitem{ABMNdata}
Country indexes are identified with MSCI Indexes.  MSCI Indexes as well as MSCI Dividend Yields,
and MSCI Price over Earning Ratios were provided by FactSet \textit{http://www.factset.com/}.
The nominal series given in local currencies have been deflated using Consumer Price Indexes available 
at the Federal Reserve Economic Data repository \textit{http://research.stlouisfed.org/fred2/}.


\bibitem{ALS1}
R.D. Arnott, F. Li and K.F. Sherrerd,
{\em Clairvoyant Value and the Value Effect}.
Journal of Portfolio Management {\bf 35} (2009), no.~3, 12--26.

\bibitem{ALS2}
R.D. Arnott, F. Li and K.F. Sherrerd,
{\em Clairvoyant Value II: The Growth/Value Cycle}.
Journal of Portfolio Management {\bf 35} (2009), no.~4, 142--157.


\bibitem{B}
{\em Thus the price of a stock will tend to move back toward its value over time. The move will often be so gradual that it is imperceptible. 
If it is fast, technical traders will perceive it and speed it up. If it is slow enough, technical traders will not be able to see it, or will be so unsure of what they see that they will not take large positions. }

{\em Still, the farther the price of a stock moves away from value, the faster it will tend to move back. 
This limits the degree to which it is likely to move away from value. 
All estimates of value are noisy, so we can never know how far away price is from value. }

{\em However, we might define an efficient market as one in which price is whithin a factor $2$ of value, i.e., the price is more
than half of value and less than twice value. 
The factor $2$ is arbitrary, of course. Intuitively, though, it seems 
reasonable to me, in the light of sources of uncertainty about value and the strength of the forces tending to cause price to return to value. 
By this definition, I think almost all markets are efficient almost all of the time. "Almost all" means
at least $90\%$.}

F. Black,
{\em Noise}.
Journal of Finance {\bf 41} (1986), no.~3, 529--543.



\bibitem{BC}
J.-P. Bouchaud and R. Cont, 
{\em A Langevin approach to stock market fluctuations and crashes}
Eur. Phys. J. B {\bf 6} (1998), 543--550. 

\bibitem{CF} J. Coakley, A.-M. Fuertes, {\em Valuation ratios and price deviations from fundamentals}, Journal of Banking and Finance {\bf 30}
(2006), 2325--2346


\bibitem{CT}
T. Cover and J. Thomas,
{\em Elements of Information Theory}, 2nd edition, 
Wiley Interscience, Hoboken, NJ, 2006.



\bibitem{DF}
It has even been suggested that the variation $\Delta P$  of an asset price  may tentatively be decomposed into three terms:
$ \Delta P = \Delta M + \Delta I + \Delta N $
where  $\Delta M$ is a slowly varying "trending" term (e.g. fractional brownian motion with Hurst exponent $>1/2$) 
related to the macroeconomic environment, 
$\Delta I$ are jumps due to the arrival of new unexpected information and 
$\Delta N$ is is a rapidly varying "mean-reverting" term (e.g. fractional brownian motion with Hurst exponent $<1/2$) 
due to "noise" traders. See, e.g., Raphael Douady's 2007 Lectures at Keio University

\bibitem{FF88} 
E. Fama and K. French, {\em Permanent and temporary components of stock prices}, 
The Journal of Political Economy, {\bf 96} no.~2 (1988), 246--273

\bibitem{J1}
E.T. Jaynes, {\em Information theory and statistical mechanics}, Physical Review. Series II {\bf 106} (4) (1957), 620--630. 

\bibitem{J2}
E.T. Jaynes, {\em Information theory and statistical mechanics II}, Physical Review. Series II {\bf 108} (2) (1957), 171--190. 

\bibitem{JT}
N. Jegadeesh, and S. Titman. {\em Returns to buying winners and selling losers: implications for stock market efficiency}, Journal of Finance, {\bf 48} (1993), no.~1, 65--91.


\bibitem{LM88}
A.W. Lo  and A.C. Mackinlay, {\em Stock market prices do not follow random walks: evidence from a simple specification test}, The Review of Financial Studies, {\bf 1} Issue 1, (1988) 41--66. 

\bibitem{NS}
C. Neri and L. Schneider,
{\em  Maximum entropy distributions inferred from option portfolios on an asset},
preprint.


\bibitem{KC}
H. Kleinert and X.J. Chen,
{\em Boltzmann distribution and market temperature},
Physica A, {\bf 383} Issue 2, 513--518.

\bibitem{S65}
P. Samuelson,  
{\em Proof that properly anticipated prices fluctuate randomly}, Industrial Management Review, {\bf 6}, (1965)
41--49. 

\bibitem{S81}
R. J. Shiller, 
{\em Do stock prices move too much to be justified by subsequent
changes in dividends?}, American Economic Review, {\bf 71} (1981), 421. 

\bibitem{Sdata} [http://www.econ.yale.edu/$\sim$shiller/data.htm]

\bibitem{Sdatadescription} According to Shiller: ``Monthly dividend and earnings data are computed from the S$\&$P four-quarter totals for the quarter since 1926, with linear interpolation to monthly figures. Dividend and earnings data before 1926 are from Cowles and associates (Common Stock Indexes, 2nd ed. [Bloomington, Ind.: Principia Press, 1939]), interpolated from annual data. Stock price data are monthly averages of daily closing prices through January 2000, the last month available as this book goes to press. The CPI-U (Consumer Price Index-All Urban Consumers) published by the U.S. Bureau of Labor Statistics begins in 1913; for years before 1913 I spliced to the CPI Warren and Pearson's price index, by multiplying it by the ratio of the indexes in January 1913. December 1999 and January 2000 values for the CPI-U are extrapolated. See George F. Warren and Frank A. Pearson, Gold and Prices (New York: John Wiley and Sons, 1935). Data are from their Table 1, pp. 11--14."

\bibitem{S00}
 R. J. Shiller, 
{\em Irrational Exuberance}, Princeton University Press, 2000.

\bibitem{WTT}
K. Watanabe, H. Takayasu and M. Takayasu,
{\em Random walker in temporally deforming higher-order potential forces observed in a financial crisis}, Phys. Rev. E {\bf 80}, 056110 (2009).


\bibitem{KE}
Usually, momentum (but we will prefer the terminology {\em quantity of motion} or {\em impetus} so as to avoid confusion with the {\em momentum effect} in finance \cite{JT}) is given by $mv$
and the kinetic energy is given by  $ K = \frac 12 m v^2.$ 

The momentum is proportional to the velocity of the price movement if many investors 
sell when price goes up and vice versa. This situation is similar to mean reverting case treated above 
and leads to a normal distribution for the returns $v$ with the variance (i.e.\ the volatility of the returns) which can be interpreted as the market temperature. 

However, one can also consider the opposite scenario: here investors are prevalently trend-followers, 
so that they prefer to invest in  {\em momentum}  stocks, i.e.\ in stocks with a recent past of high returns.
Then the momentum is
inversely proportional to the velocity and the kinetic energy will be 
$$ K = C \log |v|$$
for large $|v|$.

By the similar calculation, we have same distributions for the potential energy.


\end{thebibliography}
\end{document}